\documentclass{article}
\usepackage{algorithm}
\usepackage{algpseudocode}
\usepackage{amsmath}
\usepackage[utf8]{inputenc}
\usepackage{listings}
\usepackage{xcolor}
\usepackage{graphicx}
\usepackage{wrapfig}
\usepackage{fullpage}
\usepackage{changepage}

\usepackage{color} % Include the color package to define new colors

\definecolor{codegreen}{rgb}{0,0.6,0}
\definecolor{codegray}{rgb}{0.5,0.5,0.5}
\definecolor{codepurple}{rgb}{0.58,0,0.82}
\definecolor{backcolour}{rgb}{0.95,0.95,0.92}

\lstdefinestyle{mystyle}{
    backgroundcolor=\color{backcolour},   
    commentstyle=\color{codegreen},
    keywordstyle=\color{magenta},
    numberstyle=\tiny\color{codegray},
    stringstyle=\color{codepurple},
    basicstyle=\ttfamily\footnotesize,
    breakatwhitespace=false,         
    breaklines=true,                 
    captionpos=b,                    
    keepspaces=true,                 
    numbers=left,                    
    numbersep=5pt,                  
    showspaces=false,                
    showstringspaces=false,
    showtabs=false,                  
    tabsize=2
}

\title{\vspace{-1em}Cutting through buggy adversarial example defenses: \\ fixing 1 line of code breaks \textsc{Sabre}}
\author{Nicholas Carlini \\ \emph{Google DeepMind}}
\date{}

\begin{document}

\maketitle

\begin{abstract}
\textsc{Sabre} is a defense to adversarial examples that was accepted at IEEE S\&P 2024.
We first reveal significant flaws in the evaluation that point to
clear signs of gradient masking.
We then show the cause of this gradient masking: a bug in the original evaluation code.
By fixing a single line of code in the original repository, 
we reduce \textsc{Sabre}'s robust accuracy to 0\%.
In response to this, the authors modify the defense and introduce a new defense
component not described in the original paper.
But this fix contains a second bug; modifying one more line of code
reduces robust accuracy to \emph{below} baseline levels.
After we released the first version of our paper online,
the authors introduced another change to the defense;
by commenting out one line of code during attack we reduce the robust accuracy to 0\% again.
Then, in response to this attack, the authors released another change to the code,
reverting many changes and introducing new defense components---again;
we break \emph{this} version of the code by adding a single minus sign to
one of the attack hyperparmaeters.
\end{abstract}

\section{Introduction}

\textsc{Sabre} \cite{sabre} is a defense to adversarial examples that claims to be
$3\times$ more robust to attack than the current state-of-the-art \cite{robustbench},
bringing the best attacks on CIFAR-10 at a distortion of $\varepsilon=8/255$ from 29\% success rate down to just 11\%.
In this paper we begin with a critique of the \textsc{Sabre} paper,
and show it contains multiple evaluation flaws.
For example, the paper claims that \textsc{Sabre} is more accurate when an
adversary is attacking the defense than when not under attack, 
claims mathematically impossible levels of robustness,
and does not follow standard recommended best practices \cite{carlini2019evaluating} when evaluating adversarial 
robustness---including omitting any analysis against adaptive attacks \cite{tramer2020adaptive}.

Nevertheless, \textsc{Sabre} was accepted at IEEE S\&P 2024, one of the top computer security conferences.
(Readers may be forgiven for mistaking this defense for a \emph{different} defense---also
with a flawed evaluation---accepted at last year's IEEE S\&P 2023. \cite{carlini2023llm})

In this short paper we show that SABRE's impossible levels of accuracy can be explained
by a bug in the original evaluation; fixing the bug by changing a single line of code
increases the attack success rate increases to 100\%.
In response, in a process that will repeat three more times,
the authors make a change to the code and introduce defense components never
described in the original paper.
We show that each of these three changes contains bugs that invalidate the results we break:
the first by inserting a single function call;
the second by commenting out a single line of code;
and the third by negating one attack hyperparameter.

\section{Background \& Sabre}

We assume familiarity with evaluating the robustness of adversarial example defenses \cite{carlini2019evaluating}. 
We encourage readers not familiar with this literature to read the related work
before returning to this paper.

\textsc{Sabre} \cite{sabre} is a preprocessor defense: given an input $x$,
it first pre-processes the input with a function $p$,
and then classifies the resulting pre-processed image with a standard (adversarially trained) neural
network $f$; the resulting defended classifier is thus $f_{\textsc{Sabre}} = f \circ p$.
Because our attack is completely independent of the method by which \textsc{Sabre} works,
we do not describe it in any more detail.
Interested readers can refer to the original paper \cite{sabre} for a complete
description of the methodology.

\section{Mathematically Impossible}
\label{sec:wtf}

\textsc{Sabre} makes a number of claims that are impossible for any correct evaluation.
The impossibility of the claims it makes are sufficiently well understood that,
instead of describing the errors ourselves,
we prompt GPT-4 with a copy of the checklist in \cite{carlini2019evaluating},
a screenshot of the defense paper \cite{sabre}, and then ask
the question 
\emph{``Give one thing that is concerning about this image that suggests the paper's evaluation might be flawed. 
Focus on [some paragraph or figure].''}

\begin{itemize}
    \item \textbf{Nontrivial accuracy at $\varepsilon=0.5$.}
    ``Given the estimated value of approximately 92.66\% accuracy at a perturbation bound ($\varepsilon$) of 0.5 for the.``Ours" defense, this result appears to contradict the expectations set forth by Carlini et al. They stated that for an $\ell_\infty$ distortion of 0.5, model accuracy should not be better than random guessing because any image can be converted into a solid gray picture, making it impossible for the model to accurately predict the original content.''
    \item \textbf{Attacking the model makes it more accurate.}
    ``In the SPEECH-COMMANDS row evaluated with SABRE (the proposed defense), there is a noticeable discrepancy: the performance under the ``No Attack" condition is 83.57, which is lower than the performance under several of the adversarial attacks such as FGSM (91.21), PGD 7 steps (95.40), and PGD 20 steps (95.68). This is concerning because we would generally expect a model to perform best in the absence of any attack, as adversarial examples are specifically designed to degrade performance.

    The fact that the model performs better under attack than when there is no attack challenges the typical expectations and could suggest issues such as overfitting to the adversarial examples, a problem with the evaluation methodology, or that the reported numbers may not accurately reflect the true robustness of the model. This anomaly warrants a closer examination of the experimental setup and the robustness evaluation to ensure the validity of the results.''
\end{itemize}

\noindent
In both of these cases, the language model has correctly identified the flaws in the original paper,
and the author of this paper has checked that the model's response is accurate and true.

\section{Additional evaluation flaws}
\label{sec:wtf2}

\textsc{Sabre} also deviates from recommended practices in evaluating adversarial robustness.
Specifically, it does not follow the widely accepted best practices in evaluating adversarial robustness \cite{carlini2019evaluating} in the following ways:
\begin{itemize}
    \item  \cite{carlini2019evaluating}(\S5.2) \emph{``Verify iterative attacks perform better than single-step attacks''} In the original evaluation, the robust accuracy at 100 iterations of PGD is higher than the robust accuracy at 1 step of FGSM.
    \item  \cite{carlini2019evaluating}(\S5.4) \emph{``Verify adaptive attacks perform better than any other''} The paper performs no adaptive attack. (See the following section.)
    \item \cite{carlini2019evaluating}(\S4.4)  \emph{``Try at least one gradient-free attack and one hard-label attack''} The paper performs no gradient-free attacks or hard-label attack.
    \item \cite{carlini2019evaluating}(\S5.1)  \emph{``Compare against prior work and explain important differences''} The paper includes a comparison to PGD, but vastly under-reports the robustness of PGD as 30\% accuracy at $\varepsilon=0.3$ instead of 90\%+. 
    (It also omits comparisons to Feature Squeezing; see the following section.)
    \item \cite{carlini2019evaluating}(\S4.11)  \emph{``Attack with random noise of the correct norm''} The paper does not attempt this evaluation.
    \item \cite{carlini2019evaluating}(\S4.13)  \emph{``Perform ablation studies with combinations of defense components removed''} The paper does not do this. (See the following section.)
    \item \cite{carlini2019evaluating}(\S4.3)  \emph{``Applying many nearly-identical attacks is not useful''} The paper reports the robustness of multiple nearly-identical attacks instead of performing strong adaptive attacks.
\end{itemize}

\section{Significant flaws not observable from the paper}
\label{sec:wtf2}

We make a number of other comments on the \textsc{Sabre} defense that can not
be observed by reading the paper alone, and can only be
noticed by reading the published code or corresponding with the authors.

\paragraph{\textsc{Sabre} is not evaluated against adaptive attacks.}
An \emph{adaptive attack} is one that is ``adapted to the specific details of
the defense and attempt[s] to invalidate the robustness claims that are made'' \cite{carlini2019evaluating}
``Adaptive attacks have (rightfully) become the de facto standard for evaluating defenses
to adversarial examples.'' \cite{tramer2020adaptive}
As the authors of \cite{carlini2019evaluating} say,
``Defending against non-adaptive attacks is necessary but not sufficient. It is our firm belief
that \textbf{an evaluation against non-adaptive attacks is of very limited utility}''
(emphasis not added here, and was present in original text).

Despite the fact that the paper writes \textsc{Sabre} ``is virtually as
robust against adaptive attacks as it is against weak attacks,''
this was not actually evaluated.
The authors do not attempt an adaptive attack of \textsc{Sabre},
and in personal communication, 
the authors state that \textsc{Sabre} is not intended to be robust against future attacks---only those specific attacks considered in the paper.
When the paper writes it is robust against adaptive attacks,
the author's intent was to say that that
AutoAttack \cite{croce2020reliable}, one particular adversarial example
generation method, does not succeed more often than PGD.

But this is not what an adaptive attack means.
No attack can be said to be adaptive isolation---indeed, the phrase ``adaptive attack'' never appears in
the AutoAttack paper \cite{croce2020reliable}.
AutoAttack was not meant to be a replacement for an
adaptive attack where the adversary designs the attack
to specifically target the defense:
they even write ``we [...] see AutoAttack and adaptive attacks as complementary.
Claiming robustness to adaptive attackers without actually performing an adaptive
attack misleads the reader.

\paragraph{\textsc{Sabre} includes baselines against known-insecure defenses.}
The paper compares the proposed defense to ME-Net \cite{yang2019me}, a defense published in December 2019
that was broken two months later in February 2020 \cite{tramer2020adaptive}.
ME-Net is indeed a similar defense, in that it also pre-processes inputs in a similar
manner to \textsc{Sabre}, and so one might expect that \textsc{Sabre} would evaluate
against a similar attack strategy as the one that was used to break ME-Net,
but this was not done.

Moreover,
the authors state that they were aware that ME-Net was previously broken,
and yet (1) did not mention this fact in their paper, and (2) did not
attempt to break their defense with similar attacks to the ones that broke ME-Net.
Papers should not compare to prior insecure defenses without considering the attack that broke them,
because it is likely that the attack that broke the prior paper also will break this one.
And as we will show, the type of gradient obfuscation ME-Net suffered from is very
similar to the type of gradient obfuscation \textsc{Sabre} suffers from.

\paragraph{\textsc{Sabre} implements prior baselines incorrectly.}
The \textsc{Sabre} paper contains a comparison to adversarial training \cite{madry2017towards},
the leading methodology behind the current state-of-the-art adversarial example defenses.
Unfortunately, \textsc{Sabre} incorrectly implements adversarial training.
On MNIST, instead of adversarial training reducing attack success rate to just 11\% on
at a distortion of $\varepsilon=0.3$, the paper reports that the attack success rate against
adversarial training is over 70\%.
This makes \textsc{Sabre} appear to be a much larger improvement compared to prior work than
is actually the case.
While qualified reviewers should know the correct baseline numbers to which \textsc{Sabre} should be compared,
an under-informed reader not familiar with prior work might (incorrectly)
believe that \textsc{Sabre} has improved significantly on prior work when,
in fact, the baseline is instead incorrect.

(\textsc{Sabre} also omits any comparisons to defenses that have since improved on adversarial
training since it was published in 2018; further reducing the gap between the papers state-of-the-art
MNIST claims and the true state-of-the-art.)

\newpage
\section{Our attack}

We break Sabre three times.

\subsection{Our first break}

The paper repeatedly reports that \textsc{Sabre} is completely differentiable,
but the defense is evaluated by wrapping the preprocessing step with BPDA \cite{athalye2018obfuscated}, 
a method that replaces the gradient of the preprocessor with an estimate that $f'(x) = x$.
But \textsc{Sabre} is already differentiable; wrapping it with BDPA \emph{removes} the useful gradient
and replaces it with one that is much \emph{less} useful.
And so to break \textsc{Sabre} all we have to do is remove the unnecessary BPDA wrapping.

\definecolor{codegreen}{rgb}{0,0.6,0}
\definecolor{codered}{rgb}{0.6,0,0}

\lstdefinestyle{gitdiff}{
  basicstyle=\ttfamily\footnotesize,  % Set the basic style to small typewriter font
  escapeinside={(*@}{@*)},
  morecomment=[f][\color{codered}]-,
  morecomment=[f][\color{codegreen}]+,
  frame=single,
  rangeprefix=---\ ,
  rangesuffix=\ ---,
  includerangemarker=false
}

\begin{lstlisting}[style=gitdiff]
diff --git a/core/defenses/sabre.py b/core/defenses/sabre.py
index fe509e6..bf13629 100644          
--- a/core/defenses/sabre.py                
+++ b/core/defenses/sabre.py    
@@ -165,7 +165,7 @@ class SabreWrapper(nn.Module):
      model = Sabre(eps=eps, wave=wave, use_rand=use_rand, n_variants=n_variants)
      self.core = model                   
      self.base_model = base_model
-     self.transform = BPDAWrapper(lambda x, lambda_r: model.transform(x, lambda_r).float())
+     self.transform =            (lambda x, lambda_r: model.transform(x, lambda_r).float())
    @property                                    
    def lambda_r(self):                            
\end{lstlisting}

\noindent With this change, after running the original evaluation code, the defense achieves $<$5\% accuracy on the MNIST dataset with a perturbation bound of 0.3.
By increasing the number of attack iterations from 7 to 100, the defense accuracy drops to 0\%.
On the CIFAR-10 dataset we can achieve 0\% accuracy at a perturbation bound of
8/255 (as originally considered in the paper) with the same bugfix.

\subsection{A fix to our first break}

The authors acknowledge that our break effectively reduces \textsc{Sabre}'s robust accuracy to 0\%.
In response, the authors modify the defense and introduce a new component
that is not described in the original paper.
Specifically, the modified \textsc{Sabre} first discretizes inputs to a fixed number of digits of precision as follows:

\begin{lstlisting}[language=Python]
def precision_blend(self, x):
    if self.eps > 0:
        precision = max(min(-int(math.floor(math.log10(abs(1.25 * self.eps)))) - 1, 1), 0)
        x = self.diff_round(x, decimals=precision)
    return x

def diff_round(self, x, decimals=1):
    scale_factor = (10 ** decimals)
    x = x * scale_factor
    diff = (1 + self.error_coefficient) * x - torch.floor(x)
    x = x - diff + torch.where(diff >= 0.5, 1, 0)
    x = x / scale_factor
    return x
\end{lstlisting}

Nevertheless, the authors claim that this defense component was
intended as part of the original defense.
In personal communication with the authors,
they assert this defense component was not described in the
paper because it is not an important detail of the overall defense.
(This is despite the fact that the entire reason this component was introduced
in the first place was to prevent the attack we described above.)

In further communication with the authors,
they informed us that hyperparmaters should be set to
discretize MNIST images
so that every pixel is set to either 0 or 1 (i.e., zero decimals of precision in the code above),
and discretize CIFAR-10 images to one decimal digit of precision.

\paragraph{On differentiability.}
Throughout the paper, \textsc{Sabre} claims to be
``end-to-end differentiable to avoid gradient masking'',
claims ``innate differentiability of [their] defense mechanism'',
and claims to ``ensure end-to-end differentiability'' \cite{sabre}.
But with this new changed proposed by the authors,
this is no longer the case:
discretization is \emph{the quintessential example of nondifferentiability}.
%and omitting this fact from the paper is a critical oversight.
%
Indeed, gradient-based attacks are \textbf{only} effective when models are
usefully differentiable.
And yet, the paper states it verifies ``the absence of gradient obfuscation''.

\begin{wrapfigure}{r}{0.48\textwidth}
  \centering
  \includegraphics[width=0.48\textwidth]{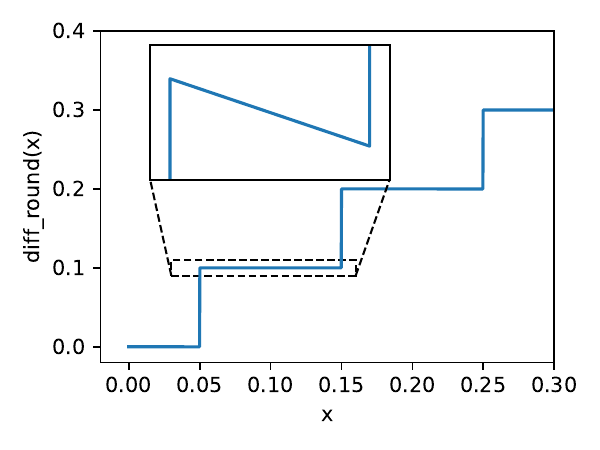}
  \vspace{-2em}
  \caption{Buggy differentiable rounding operation.}
  \label{fig:invgrad}
\end{wrapfigure}

And this is the reason why \texttt{diff\_round}  is not just a single
function call to \texttt{torch.round}: the \texttt{diff\_round} function is designed to
implement a differentiable rounding operation.
Unfortunately this function is not correct, and has a significant bug.

Figure~\ref{fig:invgrad} shows a plot of this \texttt{diff\_round} function when varying the input
from 0.0 to 0.3, and then a zoomed in view of the near-constant region.
Observe that the gradient of the function is negative \emph{almost everywhere}.
And so, therefore, when we perform gradient descent it will actually move
\emph{away} from the adversarial examples!

And so while it is technically true that this function is ``differentiable'' in the sense that $\nabla f(x)$ is defined almost everywhere,
\textsc{Sabre} contains the single most extreme form of gradient
masking the author of this paper has ever seen.
Not only does the gradient hide the true direction of the adversarial
examples, it actually points directly away from them!

\subsection{Our second break}

Previously we broke \textsc{Sabre} by removing an unnecessary BPDA wrapper.
We now insert a necessary one.

\begin{lstlisting}[style=gitdiff]
diff --git a/core/defenses/sabre.py b/core/defenses/sabre.py
index 319aebd..69fec01 100644
--- a/core/defenses/sabre.py
+++ b/core/defenses/sabre.py
@@ -61,7 +61,7 @@ class Sabre(nn.Module):
    def precision_blend(self, x):
      if self.eps > 0:
        precision = max(min(-int(math.floor(math.log10(abs(1.25 * self.eps)))) - 1, 1), 0)
-       x = self.diff_round(x, decimals=precision)
+       x = BPDAWrapper(lambda x: self.diff_round(x, decimals=precision))(x)
      return x
\end{lstlisting}

On MNIST, where the defense binarizes inputs to either 0 or 1, 
we achieve an attack success rate of 21\%.
This is significantly higher than the 13\% attack success rate of discretization alone,
as reported by \cite{xu2017feature} which was first published in 2017. % https://arxiv.org/pdf/1704.01155
Thus \textsc{Sabre} with binarization is worse than binarization alone.
In order to demonstrate conclusively that it is the binarization alone
that has caused this robustness increase, if we adjust the number
of decimals from 0 to 1 (and therefore round each input pixel to
one of $\{0, 0.1, 0.2, \dots, 0.9, 1.0\}$), then we again see that
our one-line diff again increases the attack success rate to 100\%.

On CIFAR-10, where the defense sets the number of decimals to
1 and so again discretizes each input pixel to
$\{0, 0.1, 0.2, \dots, 0.9, 1.0\}$,
we achieve an attack success rate of 100\%.
Unfortunately, this CIFAR-10 result is for a model we have trained ourselves
(but using the code published by the authors), and not for a model
trained by the authors themselves.
When we informed the authors of this second attack, they stated that this  fix
was incomplete for CIFAR-10, 
and that the defense must be adjusted differently for this model.
The following section will break this modified defense.
%However the authors have been unable to provide a description of the necessary
%changes, an implementation of the code for the fix, 
%or a pointer to where the
%necessary adjustments are described in the original paper.

\subsection{Our third break}

The first version of this paper was released on May 6th 2024, and as described above,
attacked
a model we had trained ourselves because the authors were unable to provide
a pretrained model.
Our initial paper contained the disclaimer
\emph{``When the authors provide this information we will update our paper with
an analysis of any future modifications or corrections to the currently published
defense implementation.''}
On May 24th and 27th, the authors made several corrections to the code, also released a pretrained CIFAR-10 model.
On the same day (May 27th) we broke this third release of the code,
and updated the arXiv version of this paper.

The authors' 300-LoC diff makes many changes to many pieces of the code,
again introducing changes not described in the original paper.
We omit a more complete discussion surrounding these changes from this paper,
but only remark that this is the second time the defense has been dramatically
altered since it was initially released by the authors.
We firmly believe that code for research papers should accurately reflect the experiments
actually performed when writing the paper, and should not be altered in major ways
after the fact.
But for the purpose of our paper here it is not necessary to address this topic,
because only one of the changes matters: the denoising forward pass has been updated to the following method:
\begin{lstlisting}[language=Python]
def forward(self, inputs, x0):
  b, _, m, n = inputs.shape
  c = self.in_channels

  inputs = reshape_inputs(inputs)
  x0 = reshape_inputs(x0)

  inputs = torch.cat((inputs, x0), dim=1)

  inputs = ((inputs*(v := ((1<<3)-1))).round() - ((inputs<0) & (inputs%1 != 0)).float())/v
  outputs = self.layers(inputs)

  if math.sqrt(m * n) > int(math.sqrt(m * n)):
    outputs = outputs.reshape(b, c, -1)[:, :, :m * n].reshape(b, c, m, n)
  return outputs
\end{lstlisting}

Note that line 10 of this code (again) completely discretizes the inputs---this time without any attempted smoothing.
This causes the gradient to be identically zero, and so PGD makes no progress during attack.
As a result, the defense appears to obtain 90\% robust accuracy on CIFAR-10.

To break the defense, we simply comment out this line of code when running the attack,
and then put the line back in when running the final evaluation.
This is essentially equivalent to wrapping the code with a BPDA block, but even simpler:
because the inputs are being discretized to one of eight different levels,
it is sufficient to pretend (during the attack) that there is no discretization at all.
Below is our complete break:
\begin{lstlisting}[style=gitdiff]
diff --git a/models/denoise.py b/models/denoise.py
index ba8154c..472fbec 100644
--- a/models/denoise.py
+++ b/models/denoise.py
@@ -38,7 +38,7 @@ class DenoisingCNN(nn.Module):
  inputs = torch.cat((inputs, x0), dim=1)
 
- inputs = ((inputs*(v := ((1<<3)-1))).round() - ((inputs<0) & (inputs%1 != 0)).float())/v
+ #inputs = ((inputs*(v := ((1<<3)-1))).round() - ((inputs<0) & (inputs%1 != 0)).float())/v
 
  outputs = self.layers(inputs)
\end{lstlisting}

With this single change, we reduce the defense accuracy on the paper's official
CIFAR-10 implementation with the provided CIFAR-10 model to $0\%$ robust accuracy.
As can be seen from this attack, one must be exceptionally careful when
introducing defense components that may cause gradient masking.
Each of the three releases of this paper contained a different,
but significant, gradient masking operation, despite the
paper's claims it ``ensure[s] end-to-end differentiability''.

\subsection{Our fourth break}

On June 4th the authors released a fourth version of their code in response
to our prior attack.
The purpose of this change, taken from the git commit title,
is ``refactoring with no discretization''.
That is, it it completely removes all discretization steps that were
introduced in the second and third revisions of the code.\footnote{It also
makes many other changes, again without explanation for how these changes
relate to content described in the paper. Because these other changes do
not impact our results we omit any further discussion.}
When asked if this was intentional, the authors stated that it was;
and they assert this latest version of the code accurately reflects the
experiments performed in the original paper.
At the same time, in prior communication the authors asserted on multiple
occasions that the discretization was a necessary component of their
defense that was unintentionally omitted from the first code release.

Both of these can not be true simultaneously.
The authors have now asserted that the code used in the experiments both did---and
did not---include discretization.
In this section we briefly study whether or not this version
of the code could have been the code used for the original
experiments in the paper.
Then, we show that (regardless of whether or not this code was original)
this implementation is also not robust to attack.

\paragraph{Unreproducible.}
We find that this specific version of the code can not have been what was used to perform
the original experiments: without any modifications, 
when we run the provided defense evaluation with the reference PGD attack on a
CIFAR-10 model trained by the authors we find that it \emph{increases} 
the defense accuracy to 100\%.
This is despite the original paper stating that running PGD on a CIFAR-10 trained
model should yield 93\% accuracy.
(We are unable to run the MNIST model because the authors' changes have since broken
this model and its clean accuracy is now 10\%.)

\paragraph{Our attack.}
Nevertheless,  we now break this version of the defense awaiting
any future revisions.
Our attack here is entirely trivial: whereas previously we have
\S6.1 removed a function call,
\S6.2 inserted a function call,
and \S6.3 commented out a line of code,
in this section we now insert a single minus sign.

Specifically, we make the following observation: if performing PGD stepping in
the direction of the gradient increases accuracy when under attack---why shouldn't we just
step in the opposite direction of the gradient to decrease accuracy when under attack?
To do this, all we need to do is negate the value of $\alpha$: the attack's 
step size hyperparameter.

\begin{lstlisting}[style=gitdiff]
diff --git a/notebooks/cifar10.ipynb b/notebooks/cifar10.ipynb
index 543aa90..f1e5553 100644
--- a/notebooks/cifar10.ipynb
+++ b/notebooks/cifar10.ipynb
@@ -389,7 +389,7 @@
    "metadata": {},
    "outputs": [],
    "source": [
-    "attack = EoTPGD(eps=8./255, alpha=2./255, steps=7, eot_steps=10)\n",
+    "attack = EoTPGD(eps=8./255, alpha=-2./255, steps=7, eot_steps=10)\n",
     "attack.set_model(model)\n",
     "\n",
     "printer(f'Evaluating model... [Attack=EoT-PGD 7 steps]\\n', traces_file)\n",
\end{lstlisting}

We find this attack completely breaks the latest version of the code:
negating the step direction reduces defense accuracy to 4\%, well below the random
guessing baseline.
An explanation of this phenomenon is beyond the scope of this paper.
It could be because the model was retrained (again) in a different way and this
retraining caused gradients to point backwards;
or it could be because the reconstruction model now behaves differently than
it did previously.

Given the simplicity of this attack, we aim to confirm it is effective by
providing the authors with a list of adversarial examples we generated by
negating the step size.
Despite several attempts, the authors have been unwilling to verify the efficacy
of these adversarial examples.

\section{Author Response}

We shared an advance copy of this paper describing the evaluation flaws from Sections 3-5 with the authors of \textsc{Sabre}.
They provided the following response, which we quote verbatim:

\begin{adjustwidth}{1.4cm}{1.8cm}
\subsection{Claims}
\begin{itemize}
\item Use of GPT-4: Generative models should not be used to judge research papers.
Among the many known issues of such models are their tendency to produce factually incorrect information and having their responses biased by their inputs.
\item Attacking the model makes it more accurate: The proposed framework is
primarily designed around the notion of noise reduction, such that the higher the
amount of noise observed, the stronger the filtering. Therefore, the observation that
attacking samples results in higher classification accuracies than those of benign
samples is simply indicative of the presence of noise in the benign samples. Through
the denoising of SABRE, noise is removed both from the attacked and benign
samples, allowing classifiers to be less susceptible to noise, thereby also improving
their generalization.
\item Nontrivial accuracy at $\varepsilon = 0.5$: The evaluation results reported by the paper
for an attack performed with $\varepsilon = 0.5$ assume the attacks are performed on a model
robustly trained using the SABRE defense. While the expectations set forth by
Carlini et al. are theoretically sound, this does not guarantee that any attack
method would find such perturbations.
\end{itemize}

\subsection{Evaluation}

The items below address the points made regarding the standard evaluation procedures
highlighted in the critique:
\begin{itemize}
\item (\S5.2) \emph{``Verify iterative attacks perform better than single-step attacks''}
This would
be the case if the iterative attacks only applied perturbations when the samples are
correctly classified. This is however not the case with the baseline implementations
typically used for benchmarking.

\item (\S5.4) \emph{``Verify adaptive attacks perform better than any other''}
We considered ``adaptive attacks'' to be attacks that choose the most appropriate attack per sample given
the defense, possibly out of multiple options. While the critique considers ``adaptive attacks'' to be attacks specifically designed against the defense, we believed
designing such attacks to be outside the scope of the paper and left it for future
work.

\item (\S4.4) \emph{``Try at least one gradient-free attack and one hard-label attack''} Gradient-free
and hard-label attacks were not in the context of our threat model.

\item (\S5.1) \emph{``Compare against prior work and explain important differences''}
The paper
includes comparisons against several, widely used, prior work, with parameters set
to allow fair comparison of different defenses within our evaluation settings. Unlike
the reported robust accuracy of models that have been adversarially trained with a
40-step PGD attack, our evaluations aim to assess the robustness and generalization
of models that have been adversarially trained with a 7-step PGD attack, which
explains the drop in robustness accuracy.

\item (\S4.11) \emph{``Attack with random noise of the correct norm''}
The threat model considered in the paper only considers gradient-based (whitebox) attacks.

\item (\S4.13) \emph{``Perform ablation studies with combinations of defense components removed''}
Appendix A in the paper provides ablation studies of SABRE with and
without non-core components of the proposed defense.

\item (\S4.3) \emph{``Applying many nearly-identical attacks is not useful''} The paper includes five
distinct types of adversarial attacks widely used in the literature, namely: FGSM,
PGD, EoT-PGD, CW, and AutoAttack.
\end{itemize}

\subsection{SABRE Details} 
\begin{itemize}
\item \textbf{Evaluation against adaptive attacks.} We considered ``adaptive attacks'' to
be attacks that choose the most appropriate attack per sample given the defense,
possibly out of multiple options. While the critique considers ``adaptive attacks''.
to be attacks specifically designed against the defense, we believed designing such
attacks to be outside the scope of the paper and left it for future work.

\item \textbf{Use of BPDA.} BPDA was used for the simple purpose of ensuring differentiability
regardless of the technical specificities of the implementation.

\item \textbf{Comparison against a known-insecure defense} The paper uses the ME-NET
defense as a baseline because even though it has been broken, the approach is
somewhat similar to SABRE and provides high levels of robustness when evaluated
as proposed in the original paper. The paper references ME-Net for benchmarking
and does not to suggest it as a secure standalone solution. The ME-Net defense
was broken by the AutoAttack method, which has also been used to evaluate the
robustness of SABRE.

\item \textbf{Implementation of Adversarial Training.}
The paper uses a publicly available
implementation of Adversarial Training, which is used in the literature for benchmarking. Unlike the reported robust accuracy of models that have been adversarially trained with a 40-step PGD attack, the SABRE paper assesses the robustness
and generalization of models that have been adversarially trained with a 7-step
PGD attack, explaining the drop in robustness accuracy.

\end{itemize}
\end{adjustwidth}

\subsection{Our response}

We have several disagreements with the above response, but for brevity, limit our reply to two items:

\begin{itemize}
    \item \textbf{On ``adaptive attacks'' and robustness to future attacks}.
    As we write throughout this paper, the phrase ``adaptive attack'' has a standard definition in the
    adversarial machine learning literature: they are ones that are ``adapted to the specific details of
    the defense and attempt to invalidate the robustness claims that are made'' \cite{carlini2019evaluating}.
    
    Nowhere in the original paper do the authors state their (new) definition of adaptive attacks
    (running a set of fixed attacks and reporting the minimum among these),
    or state their proposed defense is only intended to be robust against the specific attacks considered in this
    paper.
    Moreover, it would be extremely abnormal for a security paper to ever make this claim
    because any defense, once published, will be subject to scrutiny.
    If robustness to fixed attacks is all that is desired, hundreds of papers already satisfy this goal.

    \item \textbf{On threat models.} The authors claim gradient-free attacks (and attacks that just 
    add noise of the correct norm) are not ``in scope''.
    This is also incorrect: any attacker who can compute gradients can just throw away the gradient and
    only look at the loss.
    The threat model of white-box attacks encompasses the threat model of black-box attacks.
\end{itemize}

\section{Conclusion}

\iffalse
\textsc{Sabre} is the second defense to be accepted at IEEE S\&P in as many years that was
obviously incorrect and should have been immediately identified as such
from the impossible claims already present in the paper's claims.
%
But impossible claims are not the only flaws in this paper:
it makes a number of irredeemable errors that should have been
immediately clear to any qualified reviewer.
%
Last year's obviously insecure defense may be written off as a fluke---an unavoidable error
as part of the standard review process.
%
But two clearly incorrect results, accepted at back-to-back conferences, indicates
a more fundamental process failure.

Indeed, when the authors became unresponsive after being asked for a corrected
CIFAR-10 defense implementation, we asked the S\&P program chairs for assistance.
%
However the program chairs also were unresponsive to our repeated attempts 
at requests for assistance over a period of months.
%
As such we have decided to release our analysis in its current form.
\fi

\textsc{Sabre} is the second defense to be accepted at IEEE S\&P in as many years that
had clear signs of a flawed evaluation which should have been easily identified by qualified reviewers.
As we have shown, these evaluation flaws are caused by between one and three bugs in the
original evaluation that can be fixed by modifying one to three lines of code.

There is no doubt that defending against adversarial examples remains 
an interesting and important problem.
And reviewers are attracted to papers making bold claims:
it is much more tempting to accept a paper that claims near-perfect robust accuracy 
than a paper that claims a marginal improvement in accuracy.
But it is of utmost importance that papers with errors such as this
are identified during review, so as to not
suggest that incorrectly evaluating defenses is tolerable.

\paragraph{Where do we go from here?}
Adversarial machine learning is no longer a field that only studies toy problems
to develop a better scientific understanding of the robustness of machine
learning models.
There are now production systems that depend on the robustness of the underlying machine learning
models.

If, as a community, we can not tell the difference between a defense that claims
90\%+ robustness and one that actually gives 0\% robustness---on the simplest datasets
and with glaring evaluation flaws---what hope do we have for evaluating the robustness
of defenses designed to protect real, complex, production systems?

\bibliographystyle{acm}
\bibliography{main}

\end{document}